\documentclass{jfm} 

\usepackage{subfig}
\usepackage{amsmath} 
\usepackage{amssymb} 
\usepackage{array}
\usepackage{enumitem}
\usepackage{graphicx}
\usepackage{natbib} 
\usepackage{booktabs} 
\usepackage{easytable}
\usepackage{syntonly}
\usepackage{layout}
\usepackage{marginnote}
\usepackage{trackchanges}
\usepackage[font=small,labelfont=bf,tableposition=top]{caption}

\makeatletter
\newcommand{\currentfsize}{\f@size pt}
\makeatother

\renewcommand{\vec}[1]{\mathbf{#1}}

\DeclareMathSymbol{\varchi}{\mathord}{letters}{88}

\DeclareCaptionLabelFormat{andtable}{#1~#2  \&  \tablename~\thetable}

\makeatother \title
{
  Interpreting neural network models of residual scalar flux 
}

\author[Portwood and others]{
G. D. Portwood\footnote{portwood@lanl.gov}, 
B. T. Nadiga, 
J. A. Saenz,  
D. Livescu
}
\affiliation{
Los Alamos National Laboratory, Los Alamos, USA 87545\\
}

\begin{document}

\maketitle

\begin{abstract}
We show that in addition to providing effective and competitive closures, when analysed in terms of dynamics and physically-relevant diagnostics, artificial neural networks (ANNs) can be both interpretable and provide useful insights in the on-going task of developing and improving turbulence closures. In the context of large-eddy simulations (LES) of a passive scalar in homogeneous isotropic turbulence,  exact subfilter fluxes obtained by filtering direct numerical simulations (DNS) are used both to train deep ANN models as a function of filtered variables, and to optimise the coefficients of a turbulent Prandtl number LES closure.  \textit{A-priori} analysis of the subfilter scalar variance transfer rate  demonstrates that learnt ANN models out-perform optimised turbulent Prandtl number closures and Clark-type gradient models. Next, \textit{a-posteriori} solutions are obtained with each model over several integral timescales.  These experiments reveal, with single- and multi-point diagnostics, that ANN models temporally track exact resolved scalar variance with greater accuracy compared to other subfilter flux models for a given filter length scale. Finally, we interpret the artificial neural networks statistically with differential sensitivity analysis to show that the ANN models feature dynamics reminiscent of so-called ``mixed models'', where mixed models are understood as comprising both a structural and functional component. Besides enabling enhanced-accuracy LES of passive scalars henceforth, we anticipate this work to contribute to utilising  neural network models as a tool in interpretability, robustness and model discovery.
\end{abstract}

%
\section{Introduction}
The application of data-driven deep learning to physical sciences has been an emergent area of research in recent years, encouraged by the success of data-driven methods in fields such as computer vision, natural language processing, and other industrial and scientific disciplines.  While a large  variety of data-driven applications and methodologies are currently being explored in fluid mechanics \citep[][for recent reviews]{kutz17,brunton20},  the application of data-driven models, particularly artificial neural networks (ANNs), to turbulence closure has shown promise as means to perform calibration, augmentation or replacement of existing turbulence closure models \citep[e.g.][]{sarghini03,beck19,maulik19,nikolaou19,ling16,moghaddam18,salehipour19,portwood19b}.

In reduced-order frameworks such as large-eddy simulation (LES), ANNs are attractive due to their ability to (1) discover complex relationships in data, (2) effectively leverage and reduce the growing volume of high-fidelity direct numerical simulation (DNS) data and (3) be expressed algebraically such that they are tractable for mathematical analysis.  Whereas demonstrations of the first two points are valuable in assessing capabilities of ANN models, in the third point we assert that the data-driven models must be robustly interpreted before being credibly certified for engineering or scientific applications. The objective of the research presented here is to robustly interpret ANN closure models while providing statistical insight into model optimisation and performance metrics.  We conduct this analysis by considering a data-driven algebraic residual passive scalar flux model which out-performs common algebraic closures with respect to several \textit{a-priori} and \textit{a-posteriori} diagnostics.

Algebraic LES closures, the \textit{de facto} standard class of approaches to LES closure, relate resolved filter scale flow parameters to subfilter scale, or \textit{residual}, dynamics.  These models may be derived on \textbf{functional} grounds, whereby the effects of unresolved quantities on the resolved quantities are modelled, thus requiring knowledge about the nature of interactions, e.g. the net rate of kinetic energy transfer between subfilter to resolved scales.  Functional models, such as Smagorinsky-type closures \citep{smagorinsky63}, are phenomenological and limited in terms of the range of dynamics they are able to model. For instance, most Smagorinsky-type implementations are incapable of reproducing backscatter of kinetic energy from sub-filter to filter scales.  While this is a limitation in the physical sense, these characteristics lead to such models exhibiting desirable stability properties in \textit{a-posteriori} simulation.  The simplicity and numerical stability of this class of models have prompted the application of neural networks and machine learning to such functional model frameworks, for instance, in the local determination of model constants \citep{sarghini03,maulik19b,gamahara17}.                                                                      

Alternatively, algebraic models may  be developed on \textbf{structural} grounds, whereby models attempt to reconstruct dynamical quantities (e.g. the residual stress, instead of its divergence as in functional modelling), representing a broader range of residual dynamics from mathematically- or theoretically-rigorous bases \citep[][]{sagaut06}. This approach relies on the following assumptions: the structure of residual quantities are ({\it i}) universal, independent of the resolved scales, and ({\it ii}) can be determined from the resolved quantities.  For example, a class of gradient-type models are derived from the assumption of asymptotically-small filter scales \citep{clark79} and a class of scale-similarity models are developed by the imposition of scale symmetries \citep{bardina80}.  Model development from asymptotic or statistical symmetry assumptions is unreliable in practice, where such assumptions are, at best, approximate.  Frequently, these imperfect assumptions often lead to issues, such as under-dissipation of filter scale kinetic energy in gradient-type models \citep{leonard74}, such that structural models are commonly linearly combined with functional closures in so-called ``mixed'' models \citep{balarac13}. A significant contribution of neural networks in the past decade has been in optimising the balance between functional and structural components in mixed models \citep[e.g.][]{sarghini03,maulik19b,beck19}.  

By virtue of the universal approximation theorem \citep{cybenko89},
ANNs are expected to be adept at estimating closures without having to
rely on further assumptions such as those of statistical symmetry or
asymptotically-small filter scales. However, taking advantage of this
capability of ANNs to develop closures has typically led to
the characterisation of such closures as ``black-box'' models.  We contend
that appropriately designed and trained ANNs will learn the correct
structure of the closure, and that it remains to be revealed and
interpreted through proper analysis. That is, we take the point of
view that in addition to providing effective and competitive closures,
when analysed in terms of dynamics and physically-relevant
diagnostics, a learnt ANN model will be both interpretable and
provide useful insights in the on-going task of developing and
improving turbulence closures.

We explore the above point of view in the context of
statistically-stationary homogeneous isotropic turbulence coupled to a
passive scalar with a mean gradient \citep[c.f.][]{overholt96}.  
After demonstrating the capability of ANNs to learn algebraic LES closures when trained with filtered DNS data, the performance of the learnt
closures are evaluated against that obtained with optimised canonical
algebraic models in both \textit{a-priori} and \textit{a-posteriori} settings.
To the best of our knowledge, we show for the first time that a
data-driven SGS model without an assumed form, is not only stable in
\textit{a-posteriori} testing, but also out-performs canonical
structural and functional models, even after the coefficients of the
canonical models are optimised for the flows considered.  Finally,
towards interpretability of the learnt model, we perform differential
sensitivity analysis of the modelled flux with respect to input
parameters. Such an analysis permits us to (a) demonstrate that the
ANN learns a closure that is a combination of structural and
functional LES closures (cf. mixed models) and in effect (b) highlight
the potential of machine learning in accelerating model discovery and
closure development.

\section{Experimental Configuration}
\subsection{Equations of motion}
We consider flows governed by the incompressible Navier-Stokes equations,
\begin{equation}
    \frac{\partial \vec{u}}{\partial t} +
    (\vec{u} \cdot \nabla)\vec{u} 
     = -
    \frac{1}{\rho_0} \nabla p +
    \nu \nabla \cdot \nabla \vec{u}
    + A \vec{u} \text{; } \quad \nabla \cdot \vec{u} = 0
    \label{eq:nsm}
\end{equation}
where $\vec{u}=(u_x,u_y,u_z)$ is the velocity vector on the coordinate system $(x,y,z)$,  $p$ is the pressure, $\rho_0$ is the constant reference density, $\nu$ is the kinematic fluid viscosity and $A$ is a dynamic coefficient modulated by a forcing scheme, here enforcing statistical stationarity of the kinetic energy. Parametric dependence of the fluctuating quantities has been omitted for simplicity of notation.  We introduce a passive scalar with an imposed mean gradient in $z$, which is decomposed as
$\phi_t=\Phi + \phi$
where $\phi_t$ is the total scalar concentration and $\phi$ is the turbulent, spatio-temporally fluctuating  quantity with respect to the imposed mean scalar concentration $\Phi = z d\Phi/dz$.  The turbulent scalar concentration follows
\begin{equation}
    \frac{\partial \phi}{\partial t} +
    (\vec{u} \cdot \nabla)\phi 
    = - u_z \frac{d \Phi}{dz} +
    D \nabla \cdot \nabla {\phi}  \; \text{.}
    \label{eq:ts}
\end{equation}
We then apply the standard LES decomposition by applying a Gaussian filter,
\begin{equation}
G(r) = \left(\frac{6}{\pi \Delta^2}\right)^{1/2}e^{-6r^2/\Delta^2}
\label{eq:filter}
\end{equation}
where $\Delta$ is the isotropic filter length scale, to
\eqref{eq:ts} such that the filtered scalar concentration follows
\begin{equation}
    \frac{\partial \bar\phi}{\partial t} +
    (\bar{\vec{u}} \cdot \nabla)\bar\phi 
    = - \bar{u}_z \frac{d \Phi}{dz} +
    D \nabla \cdot \nabla {\bar\phi} - \nabla \cdot \vec{q}
    \label{eq:scalar_filt}
\end{equation}
where
$\bar \cdot$ is the linear filter operator ($\phi=\bar\phi + \phi'$,
$\vec{u}=\vec{\bar u}+\vec{u'}$) and $\vec{q}$ is the residual scalar
flux.  The residual flux is defined exactly as
\begin{equation}
\vec{q}_\text{DNS} \equiv \overline{\phi \vec{u}} - \bar\phi
\bar{\vec{u}} \text{ . }
\label{eq:closure}
\end{equation}

\subsection{Closure models}
We consider two canonical local closure models,
a ``functional'', Prandtl Smagorinsky (PRS) model, and a
``structural'', scalar asymptotic gradient (SAG) model.  In the PRS model,  the turbulent
diffusivity is related to a turbulent viscosity through a turbulent Prandtl number,
\begin{equation}
\mathbf{q}_\text{PRS}=-\frac{\nu_t}{Pr_t}\nabla \bar{\phi}
\label{eq:conmodel}
\end{equation}
and $\nu_t$ itself is determined using the Smagorinsky model
\begin{equation}
\nu_t = (C_s \Delta)^2 ||\bar{S}||_2 \quad. 
\end{equation}
Here $C_s$ is the Smagorinsky constant and  $||\bar{S}||_2$ is the
$L_2$ norm of the resolved strain rate tensor. While the
residual flux in this model is constrained to be parallel to the 
resolved gradient, stability properties resulting from further limiting the residual
flux to a down-gradient direction (forward-scatter) makes this a common modelling choice.

In the second SAG model, a truncated series
expansion about the mean gradient that uses the filter scale as a
small parameter \citep{clark79},
\begin{equation}
\mathbf{q}_\text{SAG}=-\frac{\Delta^2}{12}\nabla \bar{\phi} \cdot \nabla \bar{\vec{u}} \text{ ,}
\end{equation}
leaves the orientation of the residual flux unconstrained. However,
since the SAG model has been found to be under-dissipative at
filter scales sufficiently larger than dissipation scales
\citep{leonard74}, further ad-hoc fixes of the SAG model,
such as combining it with an eddy diffusivity model 
\cite[][]{clark79,balarac13}) or using an artificial `clipping'
procedure \citep{lu13} has been found to be necessary to use this model form
in LES.

As an alternative, we model the residual flux using a deep
feed-forward ANN, where such a network consists of $L$ directed and
fully connected layers 
with $N^{[l]}$  neurons in the $l^{th}$ layer, leading to a bias
vector $\vec{b}^{[l]} \in \mathbb{R}^{N^{[l]}}$ and a weight tensor
$\vec{W}^{[l]}\in \mathbb{R}^{N^{[l]}\times N^{[l-1]}}$.  The output of the
$n$-th neuron in the $l$-th layer, $a_n^{[l]}$, is given by
\begin{equation}
a_n^{[l]}= f^{[l]} \left( \sum_k^{M^{[l-1]}} W_{nk}^{[l]} a_k^{[l-1]} + b_n^{[l]}\right) 
\end{equation}
where $f$ is an activation function and $\vec{a}^{[0]},\vec{a}^{[L]} $
are the input and output objects, respectively. We optimise for $\vec{a}^{[L]}=\vec{q}_\text{ANN}$ with inputs 
$
  \vec{a}^{[0]} = \left\{
    \nabla \bar{\phi},     \nabla \bar{\vec{u}} 
\right \},
$
selected to be consistent with the fundamental assumptions of structural models outlined in the introduction and to ensure particular symmetries of the sub-filter
flux such as those of Galilean-invariance, due to  and homogeneity \citep{speziale85}.  We note that rotational-invariance, complicated by an anisotropic mean scalar gradient, is not imposed explicitly by the model or model inputs.

\subsection{Numerics and optimisation}
Numerical solution of the relevant equations are obtained by using a
standard Fourier pseudo-spectral discretisation of the equations over
a triply periodic spatial domain in conjunction with exact
pressure projection, a third-order Adams-Bashforth time integration,
and a forcing scheme as described in \citep{overholt97}.  Similar numerics to the system defined by \eqref{eq:nsm} and \eqref{eq:ts} have been extensively studied in literature \citep{overholt97,Don18,shete19} such that we do not find it necessary to discuss the numerical method in detail.  

Reference data for training and testing is obtained from DNS of \eqref{eq:ts} and \eqref{eq:nsm} using $N=512^3$ collocation points and at $Re_\lambda=170$ and $\Pr \equiv \nu / D = 1$ over 25 large-eddy times. 
For \textit{a-posteriori} model evaluation, which is described in \S 4, the filtered scalar equation \eqref{eq:scalar_filt} is solved with the three residual flux models and DNS-resolved, explicitly filtered velocity \citep[c.f.][]{vollant16}.  Initial conditions for \textit{a-posteriori} simulation are determined from explicitly filtered velocity and scalar fields obtained by the statistically-stationary DNS.

\begin{figure}
    \begin{center}
       \includegraphics{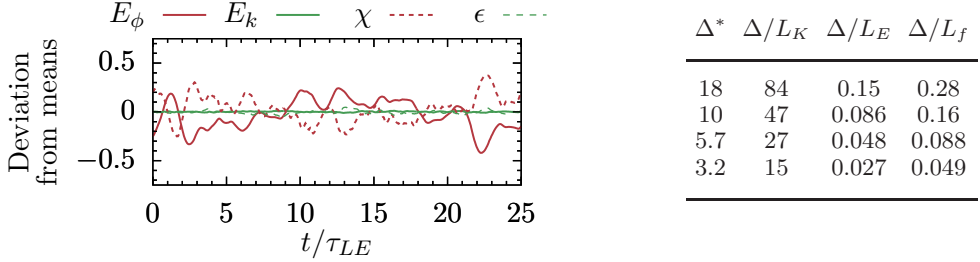}
    \hfill
        \begin{tabular}[b]{cccc}
            $\Delta^\ast$  & $\Delta/L_K$ & $\Delta/L_E$ & $\Delta/L_f$ \\
            \hline 
           18   & 84 & 0.15  & 0.28\\
           10   & 47 & 0.086 & 0.16\\ 
           5.7  & 27 & 0.048 & 0.088\\ 
           3.2  & 15 & 0.027 & 0.049\\ 
            \hline 
            \\ \\
        \end{tabular}
    \end{center}
    \captionlistentry[table]{A table beside a figure}
    \captionsetup{labelformat=andtable}
        \caption{Left, the time series of kinetic energy, scalar variance and their respective dissipation rates for the training dataset. Right, configuration of LES filter scales. A value of $\Delta^\ast=1$ indicates DNS resolution. Forcing is applied to the velocity field from largest scales in the domain until the smallest scale $L_f$. }
    \label{fig:figtab}
\end{figure}

For reference, time series of kinetic energy, scalar variance and their respective
dissipation rates 
\begin{equation*}
E_k \equiv \frac{1}{2} \langle \vec{u} \cdot \vec{u} \rangle, \quad
E_\phi \equiv \frac{1}{2} \langle \phi^2 \rangle, \quad
\epsilon \equiv \nu \langle \nabla \vec{u} : \nabla \vec{u} \rangle, \quad
\chi \equiv \kappa \langle \nabla \phi \cdot \nabla \phi \rangle, \quad 
\quad ,\end{equation*}
where the notation $\langle \cdot \rangle$ denotes a spatial average,
for the DNS solution are shown in figure \ref{fig:figtab}. The
timescales considered ensure sufficient sampling of the temporal deviations from mean statistics \citep[cf.][]{rao11,portwood19}.

We optimise the coefficients in the $\vec{q}_\text{PRS}$ model and trainable parameters in the neural network $\vec{q}_\text{ANN}$ model for four filter length scales. Optimisation is performed by using training data obtained by filtering DNS solutions with \eqref{eq:filter}, then calculating ground-truth $\vec{q}_{DNS}$ with \eqref{eq:closure}.
We note that optimisation of deterministic LES models with instantaneous \textit{a-priori} filtered quantities is not strictly consistent with the sub-filter dynamics in actual \textit{a-posteriori} simulation \citep{clark79,meneveau94b,langford99}.  While the approach will be shown to be valid in successive sections for resolved scalar dynamics of the stationary homogeneous isotropic flows considered here, we are cautious about generalising the approach to more complex flow configurations.

The selection of filter length scales are summarised in table
\ref{fig:figtab}, where $L_f$ is the outer length scale imposed by the forcing scheme, turbulent length scales are defined the standard way with
\begin{equation}
L_k = \left( \frac{\nu^3}{\epsilon} \right)^{1/4},  \quad
L_E =  \frac{E_k^{3/2}}{\epsilon}
\end{equation}
and where the filter scale is relative to the DNS grid spacing as 
\begin{equation}
\Delta^\ast\equiv\Delta/\Delta_\text{DNS} \quad .
\end{equation}
With spatio-temporal sampling at large eddy scales, the reference DNS data yielded a total of 450,000 samples, with 20\% of the data held out for out-of-sample testing. The deep feed-forward ANN consisted of 8 non-linear layers with `relu' activation, 64 neurons per layer and a final linear layer with 512 neurons before the output layer $\vec{q}$ to yield a total of 65,000 trainable parameters.  This architecture was observed to perform well with respect to different \textit{a-priori} diagnostics in a hyper-parameter study.  However, we do not dwell on other architectures
since  finding optimal architecture is not the objective of this study. Indeed, non-local approaches to neural network models have been suggested in recent literature \citep[for instance][]{pawar20,yuan20,maulik17}. Such non-local neural network models may benefit from exploiting multi-point correlations in resolved flow parameters and may be analogous to non-local mathematical models, such as deconvolutional LES models \cite[see][ for instance]{stolz99}.

\begin{figure}
    \centering
     \includegraphics{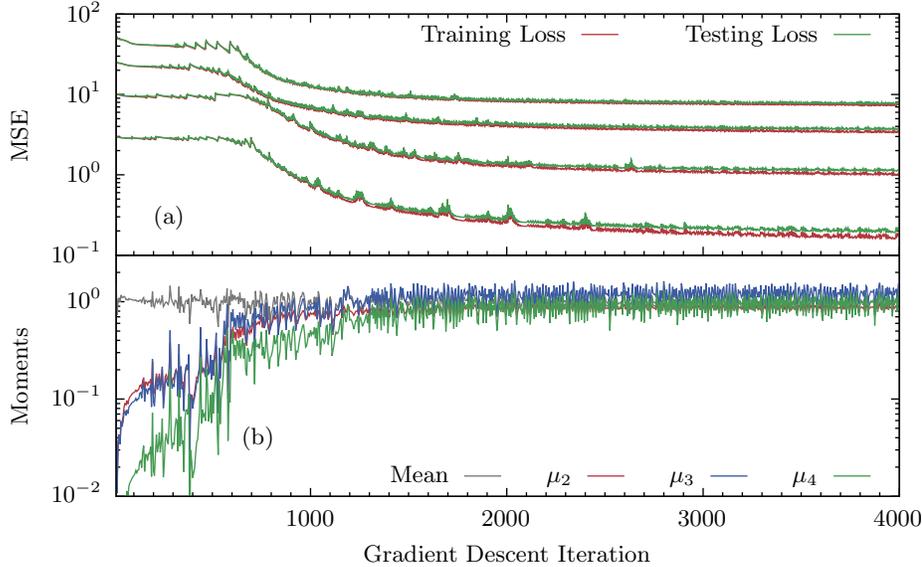}
    \caption{Training and testing loss for each filter scale as a function of gradient descent iteration is shown in panel (a), where curves with decreasing losses correspond the finer filter scales. 
    Central moments of the output parameter ${q_z}$, normalised by ground-truth data, for $\Delta^\ast=10$ as a function of gradient descent iteration is shown in panel (b). 
    Note that the mean of ${q_z}$ develops much faster than higher-order moments.
    }
    \label{fig:loss}
\end{figure}

We use the Adam optimiser to optimise the weights and biases of the $\vec{q}_\text{ANN}$ models with respect to the mean-squared-error (MSE) loss function $\mathcal{L} =E[(\vec{q} - \vec{q}_\text{DNS})^2]$. The optimiser iteratively modifies the weights and biases by using gradient descent with added momentum and dampening heuristics \citep{kingma14}.  
Training the ANN, first over 2000 iterations with a learning rate
$\alpha=10^{-3}$ and then in a second stage over a similar number of
iterations but with the learning rate decimated by an order of
magnitude, was found to be robust.  The loss function, normalised by
the variance of $\vec{q}_\text{DNS}$ for the finest filter width, is shown for the training and testing sets in figure
\ref{fig:loss}a.
We observe the loss function remaining approximately flat for each
case for the first 700 iterations. We show central moments of $q_z$,
where the $i$-th moment is defined $\mu_i$,  normalised by their
ground-truth values as a function of gradient descent iteration in
figure \ref{fig:loss}b.  During this initial training period, we
observe the mean value to converge within the first few
iterations. However, successively higher central moments demonstrate a
strong transient during this period. Most notably, $\mu_4$ does not converge until 1500 iterations. The convergence of these moments coincide with a decreasing rate-of-change of the loss function.  

\begin{figure}
    \centering
     \includegraphics{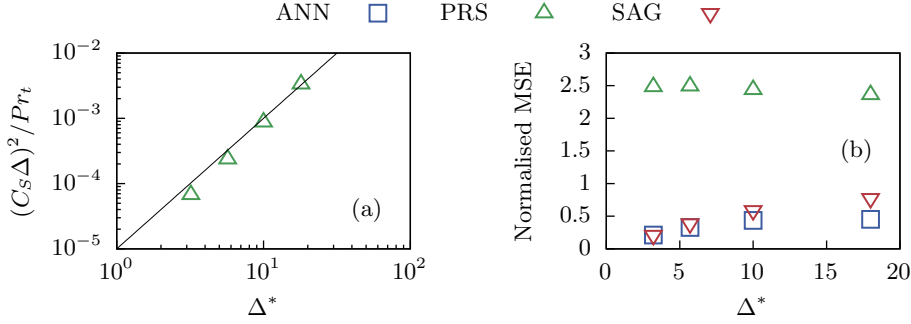}
    \caption{
    (a) Optimised coefficients for the PRS model as a function of filter width, demonstrating mild anomalous scaling from $\Delta^2$. (b) Mean-squared-error of $\vec{q}$ for each model, normalised by the variance of $\vec{q}_\text{DNS}$ for each filter width.}
    \label{fig:turbdiff}
\end{figure}

Finally, we note that in the PRS model, the constant $(C_s)^2/Pr_t$  was
similarly optimised using the same training data and loss function.
The bulk constant is shown in figure \ref{fig:turbdiff}a as a function of filter width.  We observe some slight dependence of $(C_s)^2/Pr_t$ on filter width $\Delta^\ast$.
Furthermore, the mean-squared-error of the two optimised models and the SAG model (which lacks any unknown coefficients) are shown in figure \ref{fig:turbdiff}b.  With respect to the ground-truth $\vec{q}_\text{DNS}$ variance, we observe the MSE of the PRS model to remain constant as a function of filter width.  The ANN and SAG models exhibit strong dependence on $\Delta^\ast$, with the ANN model robustly out-performing the SAG model at coarsest filter widths and an approximately linear dependence on $\Delta^\ast$ when the filter width is small.

\section{\textit{a-priori} analysis}
A useful a-priori characterisation of an LES model is its ability to
reproduce the probability distribution function (p.d.f.) of the subfilter
scale production  of
scalar variance, $\bar{P}_\phi$  as seen in the fully resolved
computations, where the production is 
given by
\begin{equation}
\bar P_\phi = \vec{q} \cdot \nabla \bar\phi \quad.
\end{equation}
The net downscale cascade of scalar variance in the setting
considered, that is, a transfer from resolved scales to unresolved
scales, leads to its mean value being positive.

\begin{figure}
    \centering
     \includegraphics{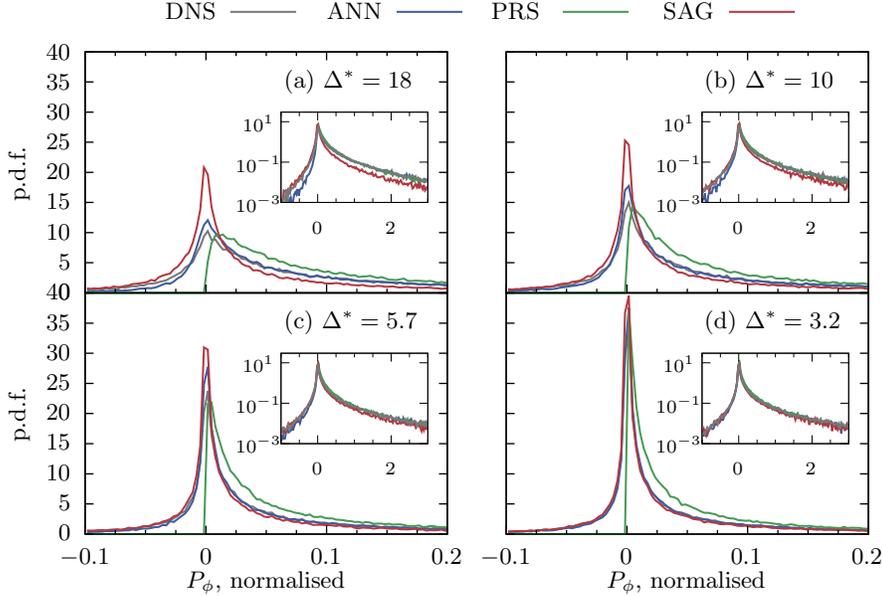}
    \caption{ \textit{A-priori} comparison of the p.d.f.s of modelled and ground-truth production, normalised by ground-truth production standard deviation, log-scales inset. 
    Note that forward- and back-scatter are more accurately captured by the ANN models. Also, note the apparent convergence of the SAG model to the ANN model and explicitly-filtered DNS results at the finest filter width in panel (d).  }
    \label{fig:apriori}
\end{figure}

The p.d.f. of the production of scalar variance is shown for the filtered DNS data and the
three closures considered at all filter sizes in figures
\ref{fig:apriori}a-d, with the computation using testing set data.
In this figure, the down-gradient nature of the residual flux in the
PRS model constrains the production to be positive-definite whereas production
in the DNS is not seen to be constrained in such a fashion. Indeed,
the SAG model is seen to be able to produce counter-gradient fluxes
(negative production)
like in the reference DNS data. The ANN model is seen to similarly produces both
counter-gradient and down-gradient fluxes.  However, as mentioned
earlier, that the SAG model tends to be under-dissipative is seen from
the fact that the integrated value $E[\bar P_\phi]$ has a relative error
of -56\% at the coarsest filter scale, and -22\% at the finest filter
scale considered. The corresponding numbers for the PRS and ANN models are
(0.19\%, 4.2\%) and (-1.5\%, -0.12\%) respectively. The ANN model is
thus seen to exhibit advantageous characteristics by accurately capturing both mean dynamics of the subfilter scalar variance transfer rate, which in unenforced in optimisation procedure, and also the distribution of the transfer rate.

\section{\textit{a-posteriori} analysis}
In an LES setting that models subfilter stresses in the momentum
equation as well (that is, in addition to the subfilter scalar flux),
dynamics of the passive scalar are affected not only by the
model for the residual scalar flux, but also by the model for the
residual momentum stress. It is easy to imagine parameter regimes
where the indirect effects of the latter dominates the direct effects
of the former, as far as the dynamics of the scalar are
concerned. Therefore, in an effort to isolate the fundamental issue of
scalar closure,  we simulate the filtered passive scalar with exact
advective coupling via explicit filtering of DNS solutions
\eqref{eq:nsm} \citep[c.f.][]{vollant16}.   

\begin{figure}
    \centering
    \includegraphics{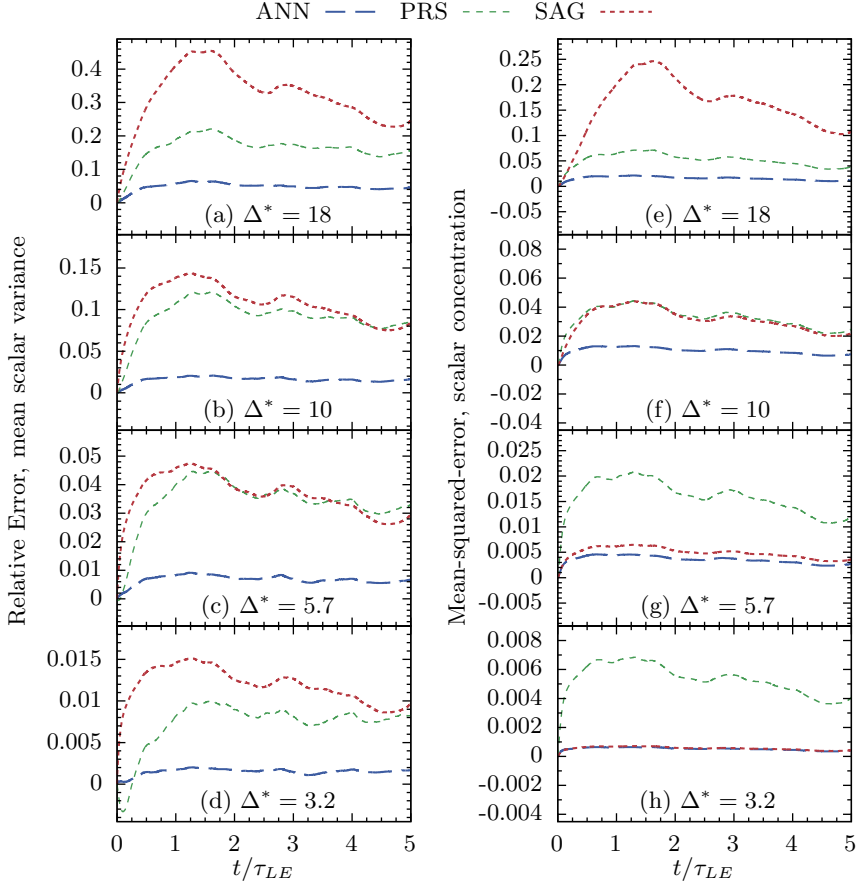}
    \caption{(a-d) Time series of relative error, with respect to filtered DNS solutions, of scalar variance $\langle \bar\phi^2 \rangle$ at multiple filter scales.  The finest filter is shown is panel (d), which demonstrates good performance for all models, while the ANN model features errors consistently an order of magnitude smaller than PRS and SAG models. Panels (e-h) indicate time series of local MSE of the scalar concentration, normalised by flux variance. Note the strong improvement of the SAG model with decreasing filter scale.}
    \label{fig:apost_time}
\end{figure}

The three closure models are implemented in an \textit{a-posteriori}
simulation fashion and run for 5 large-eddy times using explicitly filtered DNS solution, not used in testing or training datasets, as initial conditions. 
  The scalar variance for
each model, $\langle \bar\phi^2\rangle/2$, is shown in figure
\ref{fig:apost_time}a-d at four filter scales, each normalised by scalar variances obtained by explicitly filtering DNS solutions of the scalar. Visualisations of the evolution of scalars are included in Movie 1 of the accompanying supplementary material.
For the coarsest filter, shown in figure \ref{fig:apost_time}a, the SAG model
characteristically under-dissipates and scalar variance diverges from the
filtered DNS solution at early time.  The optimal PRS model initially tracks
the filtered DNS solution before $t/\tau_{LE}\approx0.1$.
For both the PRS and SAG models, the these the relative errors in resolved scalar variance soon begins to
asymptote at large values.  Whereas
the large timescale trends in the evolution of filtered scalar variance is
follow by all models, the SAG and PRS model over-predict scalar variance by 25\%
and 16\%, respectively, after 5 large-eddy turnover times.  The ANN model,
however, tracks the filtered DNS solution accurately where the scalar variance after 5 large-eddy turnover times after is within 4.7\% of the filtered DNS solution.
The same general trend are observed in figure \ref{fig:apost_time}b where
$\Delta^\ast=10$.  The accuracy of the ANN model for
$\Delta^\ast=18$ is only matched in PRS and SAG models for
$\Delta^\ast=5.7$ as shown in figure \ref{fig:apost_time}c.  When the
filter length is small, all models track the filtered DNS solution well in time
as shown in figure \ref{fig:apost_time}d as would be expected.

In addition to temporally tracking the evolution of mean scalar variance, spatio-temporally local metrics of the scalar concentration field are also important to consider for evaluation of a residual flux model.  The evolution of mean-squared-error of the turbulence passive scalar concentrations, each normalised by scalar variances obtained by explicitly filtering DNS solutions, are shown in figures \ref{fig:apost_time}e-h.  We observe broadly similar phenomenology compared to the evolution of scalar variances for PRS and ANN models across all filter widths -- with the ANN model featuring an order of magnitude smaller local error than the PRS model over time for all filter widths.  As the filter length scale decreases, we observe the SAG model to approach local errors that are very small with respect to the mean scalar variance, as is most evident in figure \ref{fig:apost_time}h.  Indeed local errors become comparable to the ANN model at the smallest filter length scale, consistent with the observations that the \textit{a-priori} subfilter scalar variance production recovers the distributions of the ground truth production at the finest filter scale for both ANN and SAG models, as previously discussed in figure \ref{fig:apriori}d.

\begin{figure}
    \centering
     \includegraphics{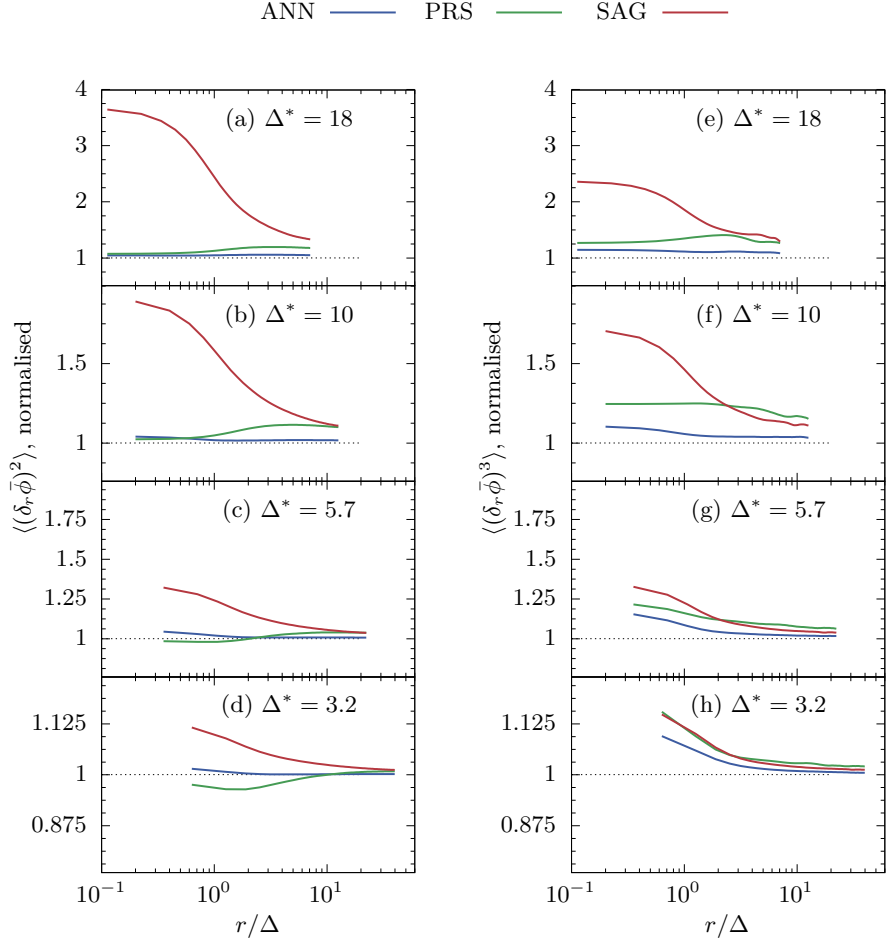}
      \caption{Two-point diagnostics. (a-d) Second-order structure functions of resolved 
      scalar in $z$, normalised by filtered DNS ground-truth data for each filter length.
      (e-h)  Third-order structure functions in $z$ for the same filters length scales, normalised by ground-truth calculations using filtered DNS.  
      }
    \label{fig:apost_adv}
\end{figure}

The temporal tracking of scalar concentration and variance is an interesting metric for
evaluation of the ANN model because the model is developed without
temporal correlations. The model also only depends on spatially-local
quantities such that reconstruction of multi-point statistics may only be
accurate if the structure of the ANN model accurately predicts the residual flux $\vec{q}$,
i.e. it does not mimic dynamics in a purely functional sense.  We evaluate two
physically relevant multi-point diagnostics in figure \ref{fig:apost_adv}.
The second-order structure function of the scalar in the direction of the mean scalar gradient, normalised
by the filtered DNS solution, after 5 large-eddy turnover times is
shown in figures \ref{fig:apost_adv}a-d for all filter scales.
For all $\Delta^\ast$, the SAG model notably misses
small-scale behaviour of the scalar, where the PRS
and ANN models exhibit more similarity with the filtered DNS results.
Except for the finest filter width, show in \ref{fig:apost_adv}d,
the PRS model begins to diverge at larger scales. The ANN model
performs comparably for both filter scales presented and the diverging behaviour of
the SAG at small scales is exaggerated with respect to the case with
$\Delta^\ast=10$.

Higher order structure functions are also instructive.  In passive
scalar turbulence forced with a mean gradient, the third-order
vertical structure function of the scalar is a physically-relevant
diagnostic which should be preserved in a physically-accurate subfilter
model \citep{warhaft00}.  Third-order scalar structure functions are
shown in figure \ref{fig:apost_adv}e-h for all filter scales, each have been normalised by the values obtained from calculations using the filtered DNS solutions.
First, we note that whereas the PRS model exhibits behaviour similar
to that of the ANN model in the second-order
structure functions, the same trends are not observed in figure
\ref{fig:apost_adv}e-h, especially at larger scales.  The SAG model
exhibits poorer consistency with large and small scales in
\ref{fig:apost_adv}e,f. Whereas the ANN model has some inconsistency
with the DNS data at small scales, these errors are smaller than as observed in other models, and it closely follows the DNS results
at larger scales.  Similar behaviour is observed for $\Delta^\ast=18$
in figure \ref{fig:apost_adv}e, except that errors may be more
exaggerated at smaller scales for SAG and PRS models. The ANN model
for this case exhibits some positive bias, almost uniformly for all
$r/\Delta$.

\section{Interpretation of data-driven models}
We propose that it is instructive to consider the sensitivity of the
residual flux to the input fields of scalar and velocity gradients as
a means to get further insight into the learnt closure. Indeed, the
automatic differentiation capability of the computational frameworks
for developing ANNs, which are necessary for the optimisation of the networks via backpropagation, can be leveraged
for such a study of the sensitivity of the residual flux to the input
parameters. This type of differential sensitivity analysis provides phenomenological interpretability of neural network mappings, but falls short of determining causality \citep{gilpin18}, as would be apparent from models obtained by symbolic approaches \citep[for instance][]{brunton16}.  

\begin{figure}
    \centering
    \includegraphics{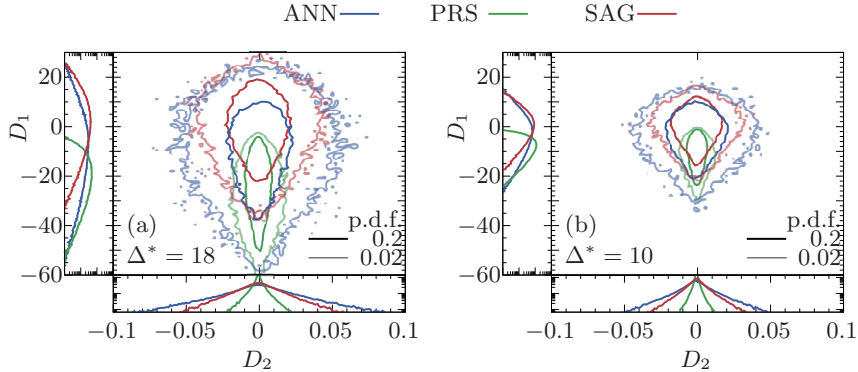}
    \caption{(a,b) joint p.d.f.s of the derivative of the vertical residual flux with respect to the vertical scalar gradient ($D_1$) and with respect to the horizontal derivative of vertical velocity ($D_2$).  Note that the distribution of the ANN and SAG models become more similar with decreasing filter scale.}
    \label{fig:interp}
\end{figure}

We compute such sensitivities with the test data
and compare them against those for the ANN, PRS and SAG models. In
particular, we present an analysis that considers
\begin{equation}
D_{1} \equiv \frac{\partial q_z}{\partial [\partial \bar \phi/ \partial z]} \; \text{ and} \;
D_{2} \equiv \frac{\partial q_z}{\partial [\partial \bar u_z/ \partial
  x]} .
\label{eq:dd}
\end{equation}
Isosurfaces of the joint pdf of \eqref{eq:dd} are shown in figure \ref{fig:interp}a for $\Delta^\ast=18$. At large $D_1$, the ANN model exhibits more similarity to the SAG model in terms of $D_2$. However, at moderate negative $D_1$, the model behaves more similar to the PRS model,  particularly in the regions near $D_2\approx 0$.  For the next finer filter scale with $\Delta^\ast=10$, as shown in figure \ref{fig:interp}b, isosurfaces of the gradients of the ANN model appear more similar to the SAG model. The interpretation of the ANN model being `intermediate' of the SAG and PRS models suggests that perhaps the ANN model may be approximated by a mixed gradient models, wherein functional gradient diffusion is added to the structural gradient model.
Furthermore, the ANN model increasingly mimics the SAG model with  decreasing $\Delta^\ast$.  The trend with $\Delta^\ast$ is consistent with the assumption of asymptotically small filter scale with the the SAG model, and suggests that the neural network learns to compensate with gradient-diffusion type dynamics as the filter scale is increased.

\section{Concluding remarks}
Encouraged by recent advancements of deep learning to industrial and technological applications, data-driven deep learning models have emerged as a promising route to calibrate, augment or replace existing models in the context of fluid turbulence.
In addition to certification of model generalisation, which was not directly considered in this study, a major criticism of such learnt
turbulence models is that they act as black-boxes, often impeding robust model certification. Responses to
this criticism have ranged from attempts to incorporate physical or
theoretical constraints, to imposing statistical symmetries,
etc. either directly in the architecture of ANNs or through the loss
function \citep[][]{raissi17}. 
Surprisingly, a point of view that has not received much
attention is that appropriately designed and trained ANNs will learn
the correct structure of the closure, and that it remains to be
interpreted and understood through proper analysis.

We have approached this point of view by the consideration of spatiotemporally local algebraic residual flux models.  By reconstructing the residual fluxes from resolved parameters, we demonstrate the capability of neural networks to discover structural relations within the data without typical assumptions of scale similarity or asymptotically small filter scale. 
In training these ANN models, we interpret the training process from a statistical perspective by demonstrating that higher-order moments of model outputs require significantly more gradient-descent iterations to converge compared to the means.
We are able to show that in addition to providing effective and competitive closures, when analysed in terms of dynamics and physically-relevant diagnostics, the learnt ANN model can indeed be interpreted in a physically and dynamically meaningful fashion.  While not determining explicit causality, such an analysis permits us to demonstrate that the ANN appears to learn a closure which features dynamics reminiscent of both structural and functional LES closures (cf. mixed models). This in effect highlights the potential of machine learning in not only providing useful insights in the on-going task of developing and improving closures, but accelerating the process of model discovery. 

\section{Acknowledgements}
The research activities of all authors are supported by a Laboratory Directed
Research \& Development (LDRD) project entitled ``Machine Learning for
Turbulence (MELT)'' (20180059DR).  
This document has been approved for unlimited release from Los Alamos National Laboratory as LA-UR-20-20405.

\bibliographystyle{jfm} 
\bibliography{bib,ml}

\end{document}